\documentclass[psfig]{kapproc} 
\usepackage{procps} 
\usepackage{psfig}
\usepackage[dvips]{graphicx}

\upperandlowercase

\let\footnote\savefootnote

\normallatexbib 

\begin{document}
\def\slash#1{#1\!\!\!/}
\articletitle{
Superconductivity with deformed Fermi surfaces
and compact stars}

\author{Armen Sedrakian}
\affil{Institute for Theoretical Physics, 
T\"{u}bingen University, D-72076 T\"{u}bingen, 
Germany
}
\email{sedrakia@tphys.physik.uni-tuebingen.de}

\begin{abstract}
I discuss the deformed Fermi surface superconductivity (DFS) 
and some of its alternatives in the context of nucleonic superfluids
and  two flavor color superconductors that may 
exist in the densest regions of compact stellar objects.
\end{abstract}


\section{Introduction}

The astrophysical motivation to study the superconducting phases of
dense matter arises from the importance of pair correlations in the 
observational manifestations of dense matter in compact stars.
If the densest regions of compact stars contain deconfined 
quark matter it must be charge neutral and in
$\beta$ equilibrium with respect to the Urca processes $d\to u
+e+\bar\nu$ and  $ u+e\to d+\nu$, where $e$, $\nu$, and $\bar\nu$ 
refer to electron, electron neutrino, and antineutrino.  
The $u$ and $d$ quarks in deconfined matter
fill two different Fermi spheres which are separated by a gap of the
order of electron chemical potential.
At high enough densities (where the typical chemical
potentials become of the order of the rest mass of a $s$ quark),
strangeness nucleation changes the equilibrium composition of the
matter via the reactions $s\to u + e+ \bar \nu$ and $u+e\to s +\nu$.  
Although the strangeness content of matter 
affects its $u$-$d$ flavor asymmetry, 
the separation of the Fermi energies remains a generic 
feature. The dense quark matter is 
expected to be a color superconductor (the early work is in 
Refs. \cite{Bar}; recent developments are summarized
in the reviews \cite{reviews1}).

Accurate description of the matter in this regime requires, first, 
tools to treat the Lagrangian of  QCD in the nonperturbative regime
and, second, an
understanding of the superconductivity under asymmetry in the
population of fermions that pair. The first principle lattice QCD
calculations are currently not feasible for the purpose of understanding the 
physics of compact stars; the effective models that are used  
rarely incorporate all aspects 
of the known phenomenology like de-confinement and chiral restoration.
Despite of the limitations of current models, a lot can be learned 
about generic features of  possible phases of dense matter 
at densities where the perturbation theory fails.
This mini-review concentrates on the second issue - the BCS
superconductors under asymmetric conditions.
Since the subject is of importance in a broader context of metallic
superconductors, nucleonic superfluids, and dilute atomic gases, and
much of our current understanding comes from the research in
these fields, I will describe the relevant physics of  non-relativistic
superconductors first (Sections 2 and 3). Section 4 discusses the 
flavor asymmetric condensates in the context of QCD using the effective
Nambu-Jona-Lasinio model; the emphasis is on the color superconducting 
state with deformed Fermi surfaces, but the discussion is sufficiently 
general to be applied to other non-BCS phases.

A comprehensive coverage of the recent developments is not possible 
in the present format; the choice of the topics will be thus personal 
and the list of the references necessarily incomplete.

\section{Homogeneous superconducting state}
  
Historically, asymmetric superconductors were studied in the early 
sixties  (shortly after the advent of the Bardeen-Cooper-Schrieffer (BCS) 
theory of superconductivity)  in the context of metallic 
superconductors with paramagnetic impurities \cite{CLOGSTON,CHANDRA,Sarma,GR}. 
There is no bulk magnetic field in these 
systems due to the Meissner effect, however the
paramagnetic impurities flip the spins of electrons in the
collisions thereby inducing an asymmetry 
in the populations of spin-up and
down fermions (which are assumed to pair in a state of total spin
zero).  The effect of impurities can be modeled by an average
spin-polarizing field which gives rise to a separation of 
the Fermi levels of the spin-up and down electrons. 
Weak coupling analysis of the BCS equations
revealed a double valued character of the gap as a
function of the difference in chemical potentials $\delta\mu \equiv 
(\mu_{\uparrow}-\mu_{\downarrow})/2$, where $\mu_{\uparrow\downarrow}$  are the 
chemical potentials of the spin up/down electrons.
The first branch corresponds  to a constant value
$\Delta (\delta\mu) = \Delta (0)$ over the asymmetry range $0\le\delta\mu \le
\Delta(0)$ and vanishes beyond the point $\delta\mu = \Delta(0)$; the
second branch exists in the range  $\Delta(0)/2\le
\delta\mu\le\Delta(0)$ and increases from zero at the lower 
limit to $\Delta(0)$ at the upper limit. Only the $\delta\mu\le
\Delta(0)/\sqrt{2}$ portion of the upper branch is 
stable (that is, only in this range of asymmetries
the superconducting state lowers the grand thermodynamic potential of
the normal state). Thus, the dependence of the superconducting state
on the shift in the Fermi surfaces is characterized by
a constant value of the gap which vanishes at the
Chandrasekhar-Clogston limit $\delta\mu_1 =\Delta(0)/\sqrt{2}$ 
\cite{CLOGSTON,CHANDRA}. The picture above, while formally
correct, is physically irrelevant to many  systems as it does not
conserve the number of particles.

Consider a BCS superconductor under the action of an external field
that produces an asymmetry in the population of the fermions; the
effect of such field is to transform the symmetric Hamiltonian ${\cal H} \to 
{\cal H}-\sigma_z I a^{\dagger}a$, where $a^{\dagger}$ and $a$ are the creation and
destruction operators (in the second quantized form), $\sigma_z$ is the
$z$ component of the vector of Pauli matrices, $I$ is the magnitude of
the asymmetry (for example, for fermions in a magnetic field 
$I=\mu_B H$, where $\mu_B$ is the Bohr magneton and
$H$ is the field intensity).
In a non-relativistic set-up the gap and the densities of spin-up 
and spin-down species are determined by the equations \cite{SAL,SedrakianLombardo,LNSSS}
\begin{eqnarray}
\label{eq1a}
\Delta_k &=& -\sum_{k'} V_{k,k'}
\frac{\Delta_{k'}}{2E_{k'}}
\left[1-f(E^{\uparrow}_{k'})-f(E^{\downarrow}_{k'})\right],\\
\label{eq1b}
\rho_{\downarrow (\uparrow)} &=& \frac{1}{2}
\sum_k\Biggl[
\left(
1+\frac{\xi_k}{E_{k}}
\right)f(E^{\uparrow(\downarrow)}_k)
+\left(
1-\frac{\xi_{k}}{E_{k}}
\right)f(-E^{\downarrow(\uparrow)}_k)
\Biggr],
\end{eqnarray}
where $V_{k,k'}$ is the pairing interaction, $f(E)$ is the Fermi
distribution function, $E_k =\sqrt{\xi^2_{k}+\Delta_{k}^2}$,  
$E_k^{\uparrow\downarrow} = E_k\pm\delta\varepsilon_k$, and the 
symmetric and anti-symmetric combinations of the single-particle
spectra are defined as 
\begin{equation}
\label{eq3}
\xi_k = \frac{1}{2}\left(\varepsilon_{k\uparrow}
+\varepsilon_{k\downarrow}\right), 
\quad \delta\varepsilon_k = \frac{1}{2} \left(\varepsilon_{k\uparrow}
-\varepsilon_{k\downarrow}\right).
\end{equation}
In the zero temperature limit the Fermi distribution 
function $f(E)$ reduces to a step function 
$\theta(-E)$. The single particle spectra
$\varepsilon_{k\uparrow(\downarrow)}$ are completely general and may
include the differences in the (effective) masses and/or
self-energies of the two species. Eqs. (\ref{eq1a}) and (\ref{eq1b}) 
should  be solved self-consistently. In the research on metallic 
superconductors these equations were decoupled and  Eq. (\ref{eq1a}) 
was solved by parametrizing the asymmetry  in terms of the difference 
in the chemical potentials, with the understanding that once the 
gap equation is solved  the densities of the constituents can be 
computed {\it a posteriori}. However, as the  value of $\delta\mu$ is 
changed so does the density of the system, i.e. such
a scheme (while being correct) does not incorporate the particle number 
conservation. If the particle number conservation is 
implemented explicitely (by solving Eqs. (\ref{eq1a}) and (\ref{eq1b})
selfconsistently) the gap becomes a single-valued function of 
the particle number asymmetry $\alpha = 
(\rho_{\uparrow}-\rho_{\downarrow})/(\rho_{\uparrow}+\rho_{\downarrow})$
\cite{SAL,SedrakianLombardo,LNSSS}.
\begin{figure}[t]
\sidebyside
{\psfig{figure=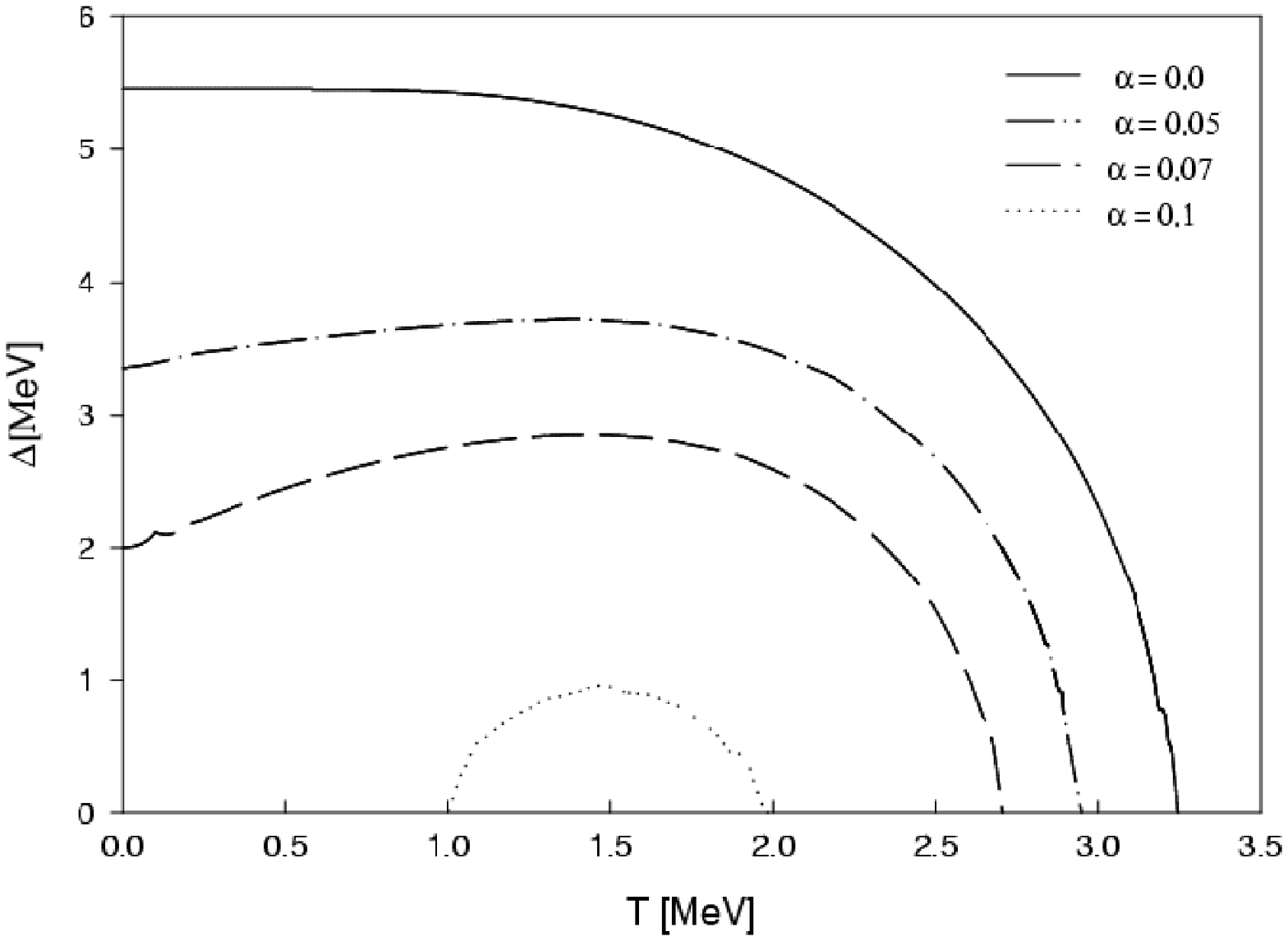,height=2.6in,width=2.3in,angle=0
}
\caption{The temperature dependence of the pairing gap for
density asymmetries $\alpha = 0.0$ (solid) 0.05 (dashed-doted) 0.07
(dashed) and 0.1 (dotted) \cite{SedrakianLombardo}.}
}
{\psfig{figure=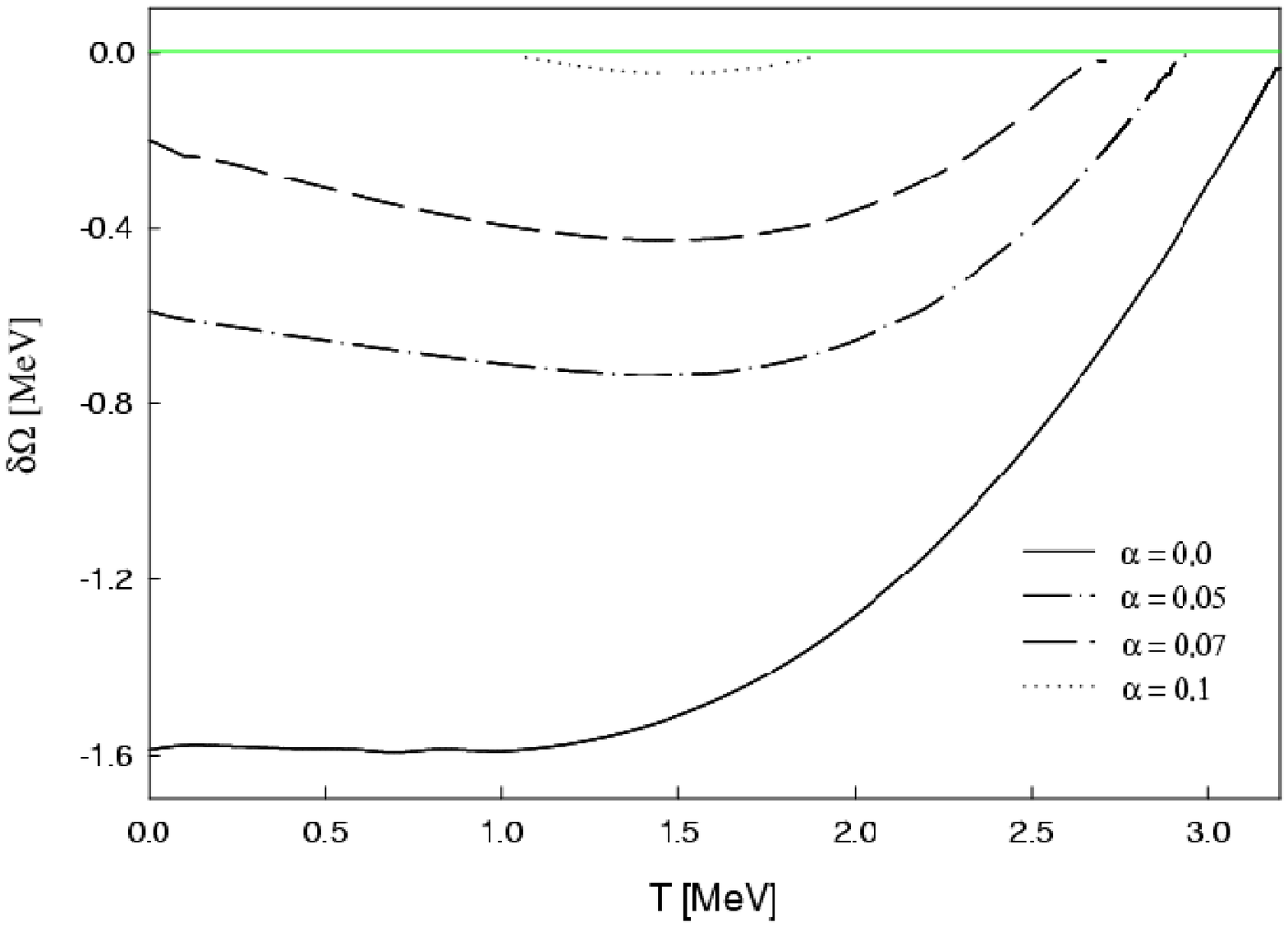,height=2.6in,width=2.3in,angle=0}
\caption{The temperature dependence of the free energy. The labeling
of asymmetries is as in Fig. 1 \cite{SedrakianLombardo}.}
}
\end{figure}
\noindent

Minimizing the free-energy of a asymmetric superconductor at fixed
density and temperature leads to stable solutions for the entire
region of density asymmetries where non-trivial solutions of
the gap equation exist \cite{SedrakianLombardo}. 
This can be seen in Figs. 1 and 2 where the
temperature and asymmetry dependence of the pairing gap and the
free-energy of a homogenous asymmetric superconductor are
shown (the examples here and in Figs. 3-8 below are
taken from the studies of the tensor $S$-wave pairing in
isospin asymmetric nuclear matter, but the overall
picture is generic to all asymmetric superconductors). In particular,
we see that for a fixed temperature the gap and the free-energy are
single valued functions of the density asymmetry $\alpha$ in a
particle number conserving scheme (contrary to the nonconserving
scheme, where double valued solutions appear).

For large asymmetries the dependence of the gap on the temperature
shows the re-entrance phenomenon - the pairing correlations are
restored (e. g. for $\alpha = 0.1$ in Fig. 1) as the temperature is
increased and a second (lower) critical temperature
appears. 
The re-entrance in the superconducting state with increasing
temperature can be attributed to the smearing of the Fermi surfaces
which increases the phase-space overlap between the quasi-particles 
that pair.
Increasing the temperature further suppresses the pairing gap 
due to the thermal excitation of the system very much the 
same way as in the symmetric superconductors. 
Clearly, the pairing gap  has a maximum at 
some intermediate temperature. The values of the two critical 
temperatures are controlled by different mechanisms:
the superconductivity is destroyed with decreasing temperature at a 
lower critical temperature when the smearing of the Fermi surfaces
becomes insufficient to maintain the phase space coherence. 
The upper critical temperature is the analog of the BCS critical
temperature and corresponds to a transition to the normal state 
because of the thermal excitation of the system. At low temperatures
the transition from the normal to the superconducting state is of the
first order, while at the temperatures near the critical temperature
- of the second order. The order of the phase transition changes
from the first to the second as the temperature is increased. Another
aspect of the asymmetric superconducting state is the gapless nature
of the excitations; in analogy to the non-ideal Bose gas where only 
part of the particles are in the zero-momentum ground state, in the 
asymmetric superconductors not all the pairs are gapped 
(see e.g. \cite{LNSSS}). The presence
of gapless excitations affects the dynamical properties of
superconductors - the heat capacity, thermal conductivity, photon 
and sound absorption, etc.

\section{Superconducting phases with broken space symmetries}


\subsection{LOFF phase}

Larkin and Ovchinnikov \cite{LO} and, independently, 
Fulde and Ferrell \cite{FF} (LOFF) discovered in 1964 that 
the superconducting state can sustain asymmetries beyond
the Chandrasekhar-Clogston limit if one pairs electrons with nonzero 
center-of-mass momentum. The weak coupling result for the
critical shift in the Fermi surfaces  for LOFF phase is $\delta\mu_2 =
0.755\Delta(0)~[>\delta\mu_1 = 0.707\Delta(0)]$. Since the condensate
wave-function depends on the center-of-mass momentum of the pair its Fourier
transform will vary in the configuration space giving rise to a
lattice structure with finite share modulus. 
This spatial variation of the order parameter in the 
configuration space implies that the condensate  
breaks both the rotational and translational symmetries.
There are thus additional massless Goldstone collective excitations
associated with the broken global symmetries (in excess of other
collective excitations that are present in the symmetric phase).

Consider again non-relativistic fermions.
Their BCS spectrum (for homogeneous systems) is isotropic; when 
the polarizing field drives apart the Fermi surfaces of spin-up and down
fermions the phase space overlap is lost,
the pair correlations are suppressed, and 
eventually disappear at the  Chandrasekhar-Clogston limit. 
The LOFF phase allows for a finite center-of-mass momentum of 
Cooper pairs $\vec Q$ and the quasiparticle spectrum is of the form
\begin{eqnarray}
\varepsilon^0_{\uparrow\downarrow}(\vec Q,\vec q) = \frac{1}{2m}\left(
\frac{\vec Q}{2}\pm\vec q\right)^2 
- \mu _{\uparrow\downarrow} + {\rm ~selfenergy ~terms}.
\end{eqnarray}
Thus, the LOFF phase requires a positive  $\propto Q^2$ increase in the 
kinetic energy of the quasiparticles which makes it less favorable
than the BCS state. However,  the anisotropic term  
$\propto \vec Q\cdot \vec q$ (which can be interpreted as a dipole
deformation of the isotropic spectrum) modifies the phase space
overlap of the fermions and promotes pairing. 
The LOFF phase becomes stable when the loss in the kinetic energy that is
needed to move the condensate is smaller than the gain in the
potential pairing energy due to an increase in the phase-space overlap.
The magnitude of the total momentum is a (variational) parameter
for a minimization of the ground state of the system.
\begin{figure}[t]
\sidebyside
{\psfig{figure=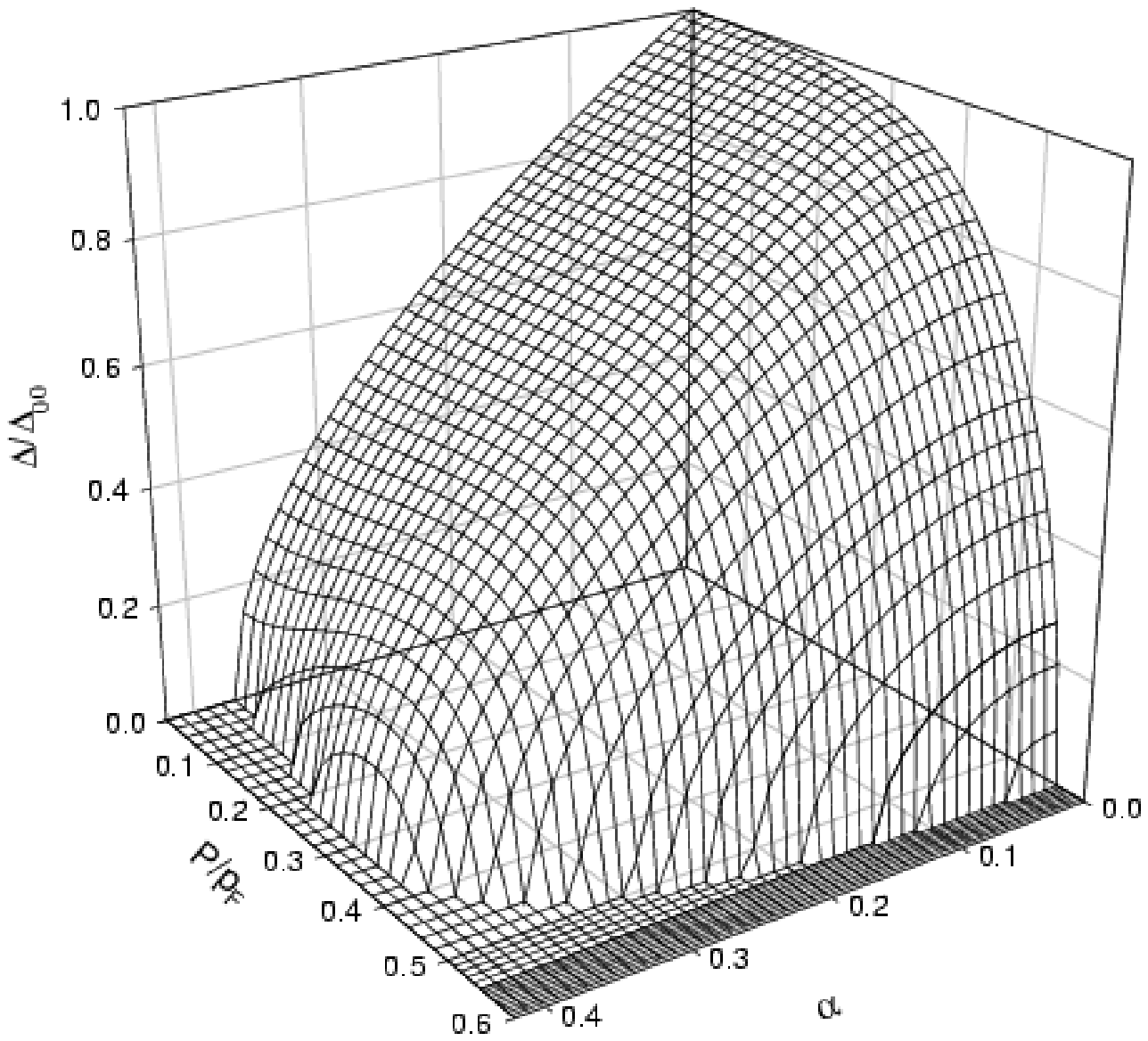,height=2.6in,width=2.3in,angle=0 }
\caption{The dependence of the pairing gap in the LOFF phase on the
density asymmetry and the total momentum of the condensate \cite{Sedrakian00}.}  }
{\psfig{figure=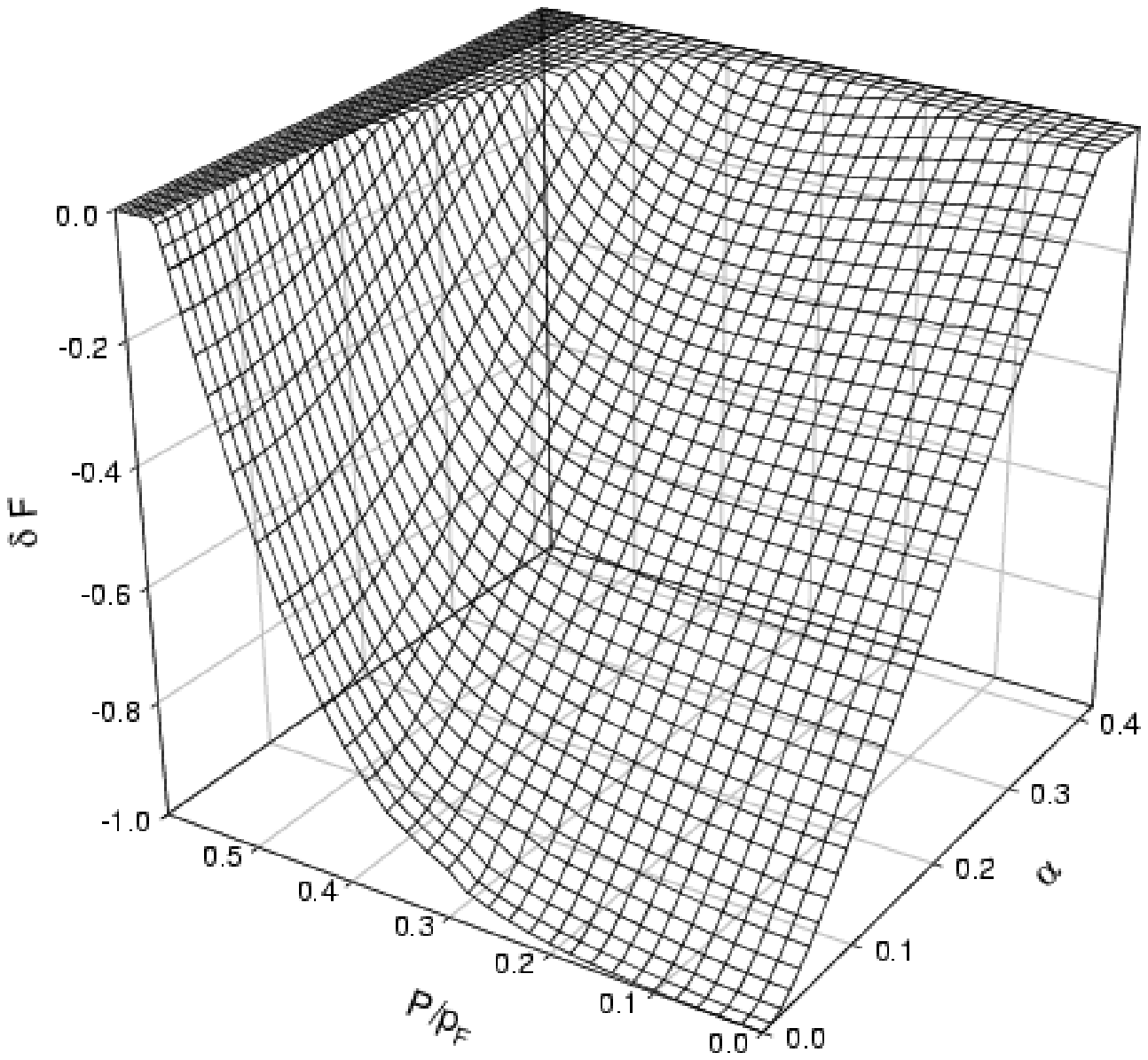,height=2.6in,width=2.3in,angle=-2.8 }
\caption{The dependence of the free energy of the LOFF phase on the
same parameters as in Fig. 3 \cite{Sedrakian00}. } }
\end{figure}
The dependence of the pairing gap and the free-energy of a LOFF 
superconductor on the total momentum of the condensate and the density 
asymmetry is shown in Figs. 3 and 4 \cite{Sedrakian00}. 
The self-consistent solution of Eqs. (\ref{eq1a}) and
(\ref{eq1b}) leads to a single valued pairing gap and stable
superconducting state for arbitrary finite momentum of the condensate, 
in particular $Q\to 0$ limit is consistent with the earlier discussion
of homogeneous asymmetric BCS condensates. For large enough asymmetries the minimum
of the free-energy moves from the $Q=0$ line to intermediate values of
$Q$, i.e., the ground state of the system corresponds to a condensate
with nonzero center-of-mass momentum of Cooper pairs. 
Note that for the near critical range of asymmetries the
condensate exists only in the LOFF state and its dependence on the
total momentum shows the re-entrance behavior seen in the temperature
dependence of the homogeneous superconductors. Clearly, a single wave-vector
condensate is an approximation; in general the LOFF phase can acquire
a complicated  lattice structure. A large number of lattice 
structures were studied in Refs. \cite{BR,Bowers} in the Ginzburg
Landau regime, were it was found that the face-centered cubic lattice has 
the lowest energy. The LOFF phase obtains additional collective 
excitations (Goldstone modes) due to the breaking of the  
rotational and translational continuous space symmetries 
\cite{Casalbuoni:2002pa}. Identifying the 
order of the phase transition from the LOFF to the normal state is a
complex problem and depends, among other things, on the preferred 
lattice structure (see Ref. \cite{Casalbuoni:2003jn} and references therein).

\subsection{DFS phase}

To motivate our next step recall 
that  the LOFF spectrum can be view as a dipole [$\propto P_1(x)$]
perturbation of the spherically symmetrical BCS spectrum,  
where $P_l(x)$ are the Legendre polynomials, and $x$ is the
cosine of the angle between the particle momentum and the total
momentum of the Cooper pair. The $l=1$ term in the expansion about
the spherically symmetric form of Fermi surface corresponds
to a translation of the whole system, therefore it preserves the
spherical shapes of the Fermi surfaces.
We now relax the assumption that the Fermi surfaces are spherical
and describe their deformations by expanding the spectrum in spherical 
harmonics \cite{DFS1,DFS2}
\begin{eqnarray}
\label{exp}
\varepsilon_{\uparrow(\downarrow)} 
(\vec Q, \vec q)= \varepsilon_{\uparrow(\downarrow)}^0 (\vec Q, \vec q)
+\sum_l \epsilon_{l,\uparrow(\downarrow)}P_l(x) ,
\end{eqnarray}
where the coefficients $\epsilon_l$ for $l\ge 2$ describe the
deformation of the Fermi surfaces which break the 
rotational O(3) symmetry down to O(2). The O(2) symmetry axis 
is chosen spontaneously and clearly need not coincide with the
direction of the total momentum (this subsection assumes $Q=0$). 
\begin{figure}[b]
\sidebyside
{\psfig{figure=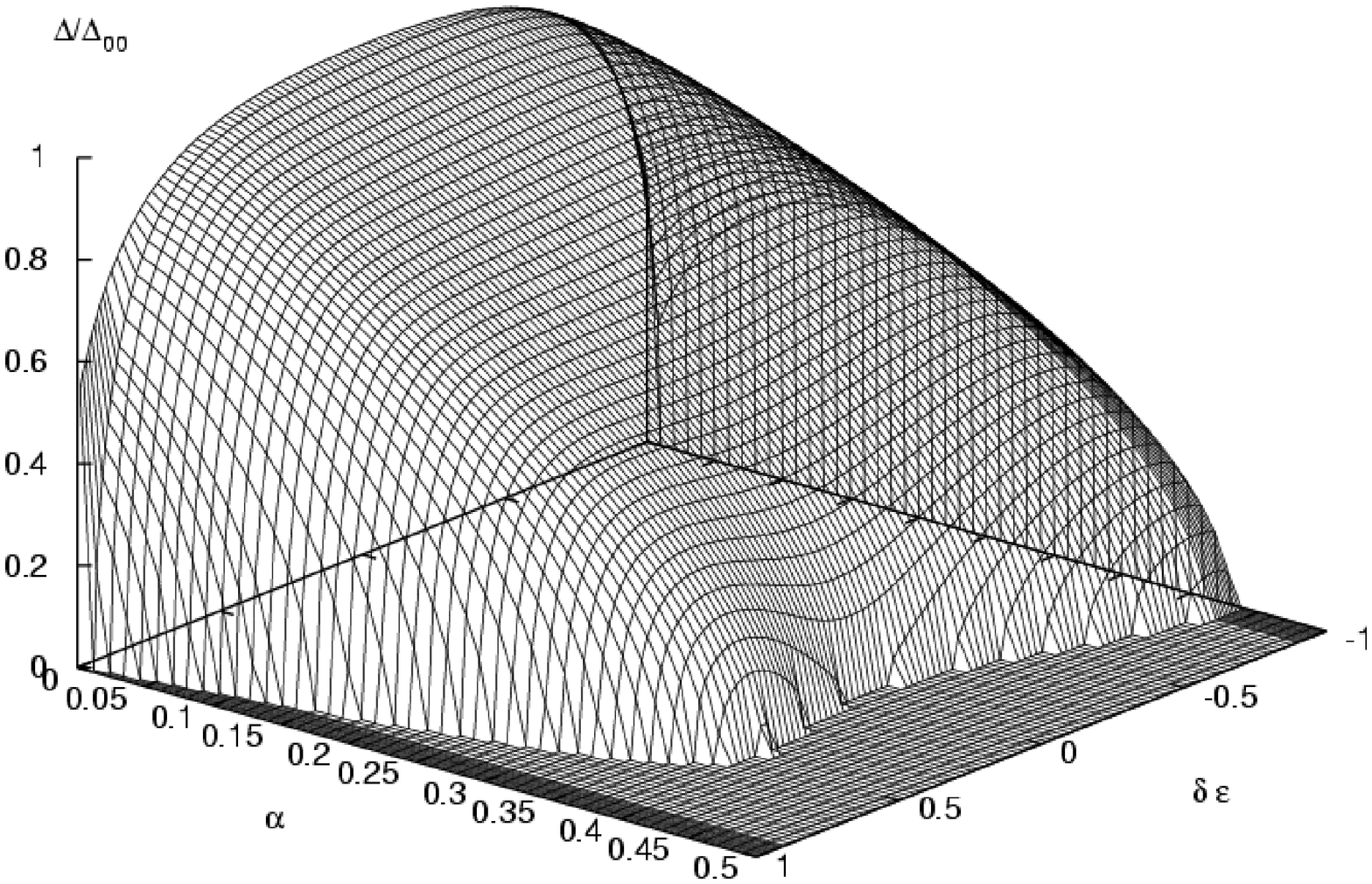,height=2.6in,width=2.3in,angle=-90
}
\caption{The dependence of the pairing gap in the DFS phase
on the density asymmetry  and the total momentum of 
the condensate \cite{DFS1}.}
}
{\psfig{figure=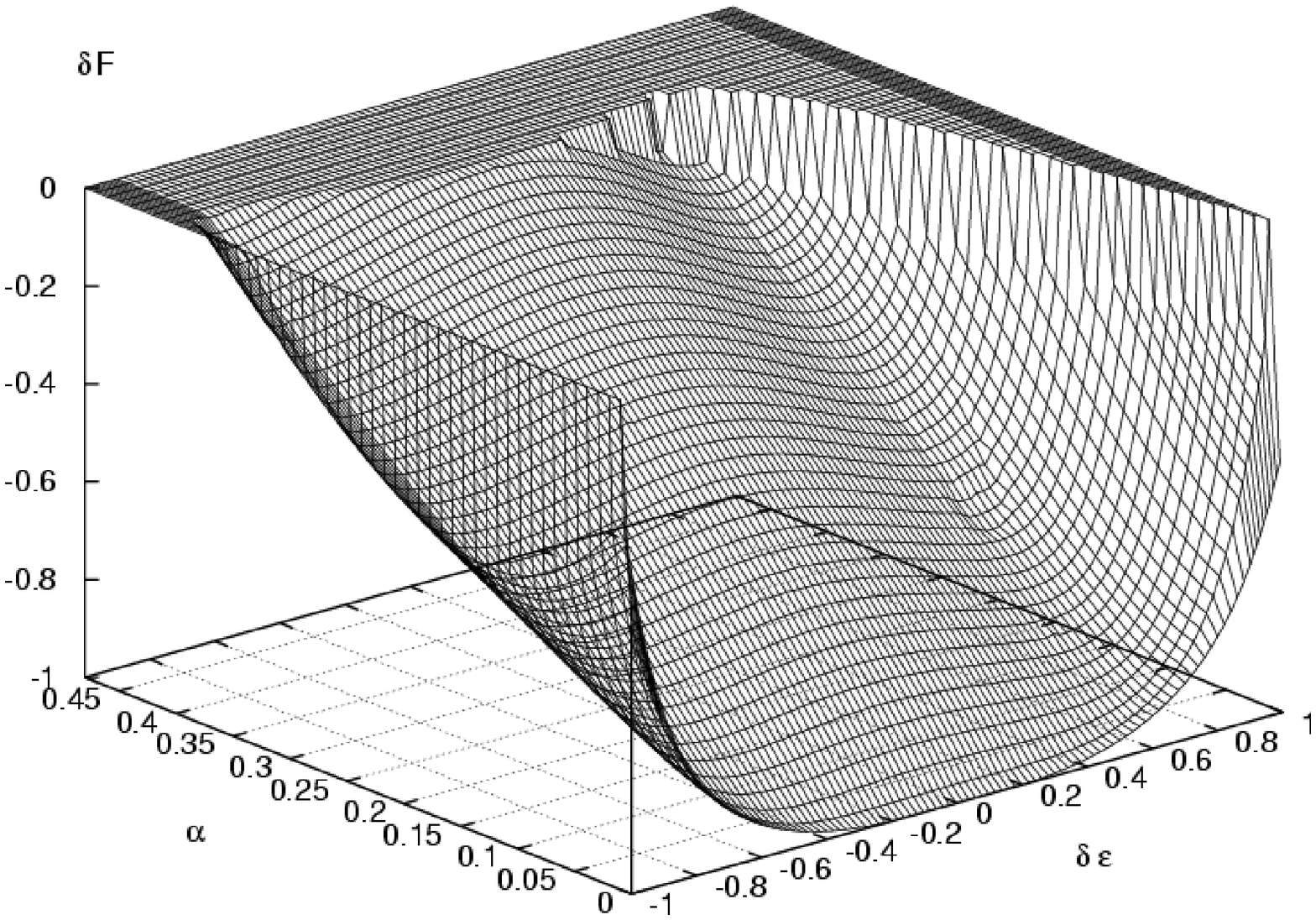,height=2.6in,width=2.3in,angle=-90
}
\caption{The dependence of the free energy of the DFS phase on the
same parameters as in Fig. 5 \cite{DFS1}.}
}
\end{figure}
A single-component and spatially homogeneous system of non-interacting 
particles fills the states within its Fermi sphere homogeneously. 
In Fermi liquids the homogeneous filling prescription is 
extrapolated to (arbitrarily strongly) interacting quasiparticles. It is by no
means obvious that such a prescription should remain valid for two or 
multi-component systems which interact, for example, by pairing
forces. The expansion (\ref{exp}) is an example of a non-Fermi-liquid 
prescription for filling the particle states within a volume bounded by a (deformed) 
Fermi surface; the deformations are stable if they lower the ground state energy 
of the system with respect to the undeformed state. Note that in
solids the Fermi surfaces are rarely spherical while their topology is
dictated by the form of and interactions with the ion lattice. 
Note also that one should distinguish between the spontaneous
deformation of Fermi-surfaces and explicit breaking of rotational
symmetry by external fields. In the latter case the initial 
Lagrangian contains  term(s) that explicitely  break the symmetry
and the resulting anisotropy of the self-energies 
can be interpreted as  a deformation of the  
Fermi surfaces~\cite{Nakano:2003rd}.  These type of deformations
are interaction induced and  are unrelated to the spontaneous 
deformations that appear even if the interaction is O(3) symmetric.
\begin{figure}[t]
\includegraphics[height=3.5in,width=3.5in,angle=-90]{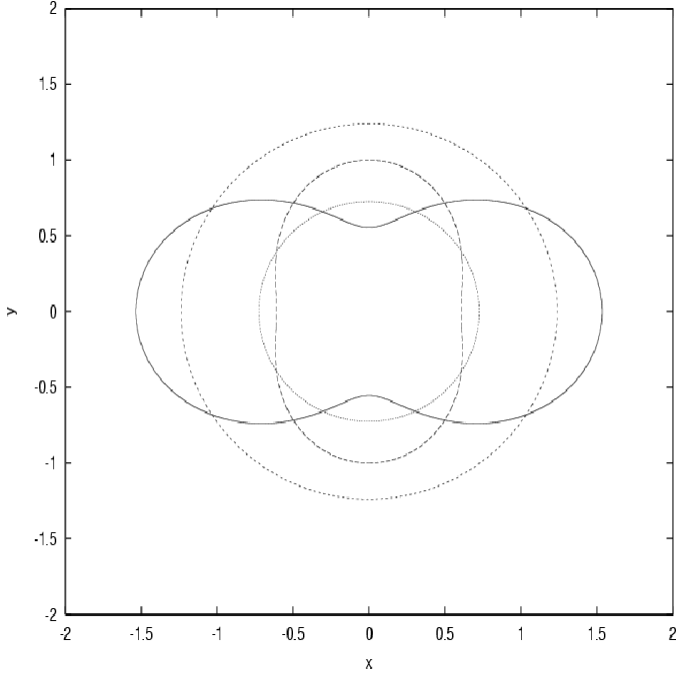}
{
\caption{
A projection of the Fermi surfaces on a plane parallel to the axis of 
the symmetry breaking. The concentric circles correspond to the two 
populations of spin/isospin-up and down fermions in spherically
symmetric state ($\delta\epsilon  = 0$), while the  deformed figures 
correspond to the state with  relative deformation 
$\delta\epsilon = 0.64$. The density asymmetry is $\alpha = 0.35$.}
}
\end{figure}

In practice, the deformation 
parameters $\epsilon_l$ ($l\ge 2$) are determined from the minimization of 
the free-energy of the system in full analogy to the total
momentum $Q$ of a Cooper pair. And they can be determined in a volume
conserving manner by solving Eqs. (\ref{eq1a}) and (\ref{eq1b})
simultaneously (the resulting phase is abbreviated as the DFS
phase).
It is convenient to work with dimensionless deformation parameters
corresponding to relative and conformal deformations defined as
$\delta\epsilon = (\epsilon_{2,\uparrow}-\epsilon_{2,\downarrow})/2\mu$ and 
$\epsilon = (\epsilon_{2,\uparrow}+\epsilon_{2,\downarrow})/2\mu$,
where $\mu$ is the chemical potential in the symmetric phase.
The dependence of the pairing gap and the free-energy of the DFS phase
on asymmetry and the relative deformation (at zero conformal
deformation) is shown in Fig. 5 and 6, respectively.
Although the density asymmetry ($\alpha$) changes in the interval
$[-1;1]$ in general,  the symmetry of the equations with respect 
to the indices labeling the species reduces the range of $\alpha$  to
$[0;1]$. The relative deformation is not bounded and can assume both 
positive and negative values. Fig. 7 shows a typical configuration of
deformed Fermi surface which lowers the ground state energy below the
non-deformed state. For $\alpha = 0$ Eqs. (\ref{eq1a}), (\ref{eq1b}) and 
(\ref{eq3}) are symmetrical under interchange of the sign of $\delta\epsilon$ and
the critical deformation for which the pairing vanishes is the same  
for prolate/oblate deformations.
For finite $\alpha$  and  the positive range of $\delta\epsilon$, 
the maximum value of the gap is attained at constant 
$\delta\epsilon$; for negative $\delta\epsilon$ the maximum 
increases as a function of the deformation and saturates around 
$\delta\epsilon \simeq 1$. As for the LOFF phase, the re-entrance
phenomenon sets in for large asymmetries as $\delta\epsilon$ is
increased from zero to finite values. And 
the mechanism by which the superconductivity is revived is based on
the same phase-space argument, but involves a deformation of the 
Fermi surfaces rather than a motion of the condensate.
Unlike the LOFF phase, the  DFS phases does not break the 
translational symmetry of a superconductor (there are still additional 
collective excitations generated by the broken continuous rotational 
symmetry). 

\subsection{DFS vs LOFF}

Which patterns of symmetry breaking  are the most 
favorable if the Cooper pairs move with a finite center-of-mass 
momentum and the Fermi surfaces are allowed to be deformed? 
To answer this question we use the set-up 
of the previous sections and choose to work with the spectrum
(\ref{exp}) at finite values of $Q$ and $\delta\epsilon$ \cite{DFS2}.
Figure 8 displays the difference between the free energies of the
superconducting and normal states $\delta{\cal F}$ normalized to its
value in the asymmetric BCS state 
$\delta{\cal F}_{00}=\delta{\cal F}(Q=0, \delta\epsilon = 0)$. Since the
energy of the pair interactions scales as the square of the pairing
gap, the shape of the $\delta{\cal F}$ surface closely resembles 
that of the pairing gap (see for details Ref. \cite{DFS2}). 
The asymmetric BCS state is the stable ground state of the system 
($\delta{\cal F} < 0$), however it corresponds to a saddle point - 
perturbations for finite $\delta\epsilon$ and $Q$ are unstable 
towards evolution to lower energy states. For the pure LOFF phase
($\delta\epsilon = 0$)  the ground state corresponds to finite momentum $Q\sim
0.5$ (in units of Fermi-momentum $p_F$).  For the pure DFS phase ($Q = 0$) there are two
minima corresponding to $\delta\epsilon \simeq -0.8$ and 
$\delta\epsilon\simeq 0.55$, i.e., prolate and oblate deformations of
the minority and majority Fermi spheres, respectively.  
\begin{figure}[t]
{\psfig{figure=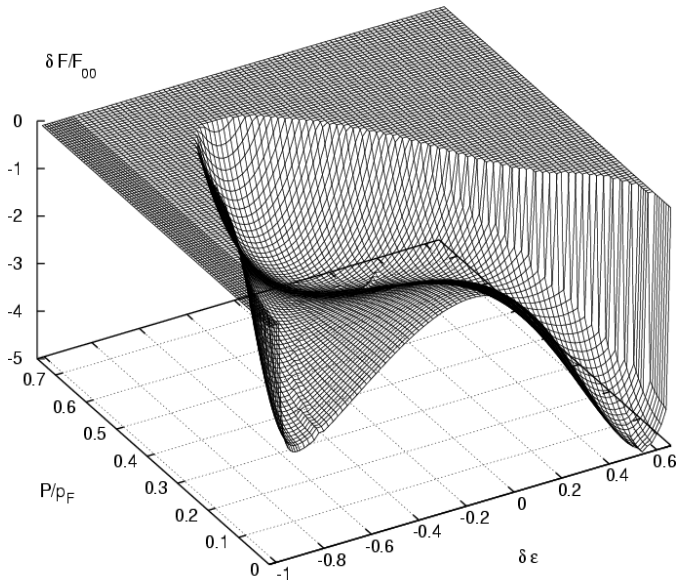,height=3.0in,width=3.5in,angle=-90
}
\caption{The dependence of the free energy of the combined DFS and
LOFF phases on the center-of-mass momentum of the pairs $Q$ (in units
of Fermi momentum $p_F$) and the relative deformations
$\delta\epsilon$ for a fixed density asymmetry \cite{DFS2}.}
}
\end{figure}
In  general the position of the minimum of $\delta{\cal F}$ in the
$\delta\epsilon$-$Q$ plane (passing through the minima of the
limiting cases) prefers either large deformations or large finite
momenta. The absolute minimum energy state corresponds to
$\delta\epsilon\simeq 0.55$ {\it and } $Q = 0$; that is, while the
LOFF phase is a local minimum state, it is unstable towards
evolution to a pure DFS phase with oblate and prolate deformations of
the majority and minority Fermi spheres.  Further work will
be needed to clarify how universal are these features. 
In particular, the assumption of a single wave-vector LOFF phase
should be relaxed.

\subsection{Alternatives}

To complete our discussion of non-relativistic superfluids let us 
briefly mention some of the alternatives to the LOFF and DFS phases.
One possibility is that the system prefers a phase 
separation of the superconducting and normal phases in real space,
such that  the superconducting phase contains particles with the same 
chemical potentials, i.e. is symmetric, while the normal phase
remains asymmetric \cite{Bedaque:1999nu,BCR}. 

Equal spin (isospin, flavor) pairing is another option, 
if the interaction between the same spin particles is
attractive \cite{Schaefer00,Buballa:2002wy,Alford1}. 
Since the separation of the Fermi surfaces 
does not affect the spin-1 pairing on each Fermi surface,
an asymmetric superconductor evolves into a spin-1 superconducting 
state (rather than a non-superconducting state) as the 
asymmetry is increased. Therefore, the spin-1 pairing is the limiting 
state for very large asymmetries.
If the states corresponding to
different Fermi surfaces are characterized by spin (as is the case
in the metallic superconductors) the pairing interaction in a
spin-1 state should be $P$ wave and  the transition is from 
the $S$ to the $P$ wave pairing.  For larger number of discrete quantum
numbers that characterize the fermions (say spin and isospin) 
the transition may occur between different $S$ wave phases 
(e.g. from isospin singlet to the isospin triplet state in 
nuclear matter).

\section{Flavor asymmetric quark condensates}

This section deals with the color superconductors and describes a
straightforward formalism for extending the discussion of the previous
sections to relativistic systems. Below it will be assumed that the
superconducting phase is chirally symmetric and particles are
interacting only via a pairing force (self-energy and vertex
renormalization are ignored).
The flavor asymmetric color superconducting quark matter appears in
the context of the two-flavor pairing (2SC-phase) described by the 
order parameter \cite{Bar,reviews1}
\begin{eqnarray}\label{QCDeq2} 
\Delta \propto \langle \psi^T(x)C\gamma_5\tau_2\lambda_2\psi(x)\rangle ,
\end{eqnarray}
where $C=i\gamma^2\gamma^0$ is the charge conjugation operator,
$\tau_2$ is the second component of the Pauli matrix acting in the
SU$(2)_f$ flavor space, $\lambda_A$ is the antisymmetric Gell-Mann
matrix acting in the  SU$(3)_c$ color space.
The Ansatz for the order parameter implies that the color
SU$(3)_c$ symmetry is reduced to SU$(2)_c$ since only two of the quark 
colors are involved in the pairing while the third color remains
unpaired. The effective Lagrangian density of the Nambu-Jona-Lasinio
model  that describes our system is of the form 
\begin{eqnarray}\label{QCDeq1}
{\cal L}_{\rm eff} &=& \bar \psi (x)
(i\gamma^{\mu}\partial_{\mu})\psi(x)\nonumber\\
&+&G_1(\psi^TC\gamma_5\tau_2\lambda_A\psi(x))^{\dagger}
(\psi^TC\gamma_5\tau_2\lambda_A\psi(x)).
\end{eqnarray}
The partial densities and the gap equation for  the up and down paired 
quarks can be found from the fixed points of the thermodynamic 
potential density $\Omega$ 
\begin{eqnarray}\label{QCDeq3} 
\frac{\partial\Omega}{\partial\Delta} = 0, \quad
-\frac{\partial\Omega}{\partial\mu_f} = \rho_f;
\end{eqnarray}
the flavor index $f=u,d$ refers to up ($u$) and down ($d$) quarks.
For the Lagrangian density defined by Eq. (\ref{QCDeq1}) and the pairing 
channel Ansatz  (\ref{QCDeq2}), the finite temperature thermodynamical 
potential $\Omega$ per unit volume is
\begin{eqnarray}\label{eq4}
\Omega= -2 \sum_{p,ij}
\Biggl\{2p +\frac{1}{\beta}{\rm log}\left[f(\xi_{ij})\right]^{-1}
+E_{ij}+\frac{2}{\beta}{\rm log}\left[f(s_{ij} E_{ij})\right]^{-1}
\Biggr\}+\frac{\Delta^2}{4G_1},
\end{eqnarray}
where the indeces $i,j = (+,-)$ sum over the four branches of the 
paired and unpaired quasiparticle spectra defined, respectively, as 
$\xi_{\pm\pm} = {(p\pm\mu)}\pm\delta\mu$
and 
$
E_{\pm\pm} = \sqrt{(p\pm\mu)^2+\vert \Delta\vert^2}\pm\delta\mu,
$
where $\delta\mu = (\mu_u-\mu_d)/2$ and $\mu = (\mu_u+\mu_d)/2$
with $\mu_u$ and $\mu_d$ being the chemical potentials of the up and 
down quarks; $s_{+j} =1 $ and $s_{-j}={\rm sgn}(p-\mu)$ and
$f(\xi_{ij})$ are the Fermi distribution functions.
The variations of the thermodynamic potential (\ref{eq4}) provide 
the gap equation
\begin{eqnarray} \label{eq5a}
\Delta &=& 8G_1\sum_p\Bigg\{
\frac{\Delta}{E_{+-}+E_{++}}
\left[
{\rm tanh}\left(\frac{\beta E_{++}}{2}\right)
+{\rm tanh}\left(\frac{\beta E_{+-}}{2}\right)\right]+
{\rm ex}\Bigg\},\nonumber\\
\end{eqnarray}
where ex abbreviates a second term which follows from the first one via a
simultaneous interchange of the signs; the partial densities of the 
up/down quarks are 
\begin{eqnarray} \label{eq5b}
\rho_{u/d} &=& \sum_{\vec p,j=\pm}\Bigg[2f(\xi_{-\mp})-2f(\xi_{+\pm})
\mp\left(1\pm\frac{\xi_{j-}+\xi_{j+}}{E_{j-}+E_{j+}}\right)
{\rm tanh}\left(\frac{\beta E_{j-}}{2} \right)
\nonumber\\
&&\pm\left(1\mp\frac{\xi_{j-}+\xi_{j+}}{E_{--}+E_{j+}}\right)
{\rm tanh}\left(\frac{\beta E_{j+}}{2} \right)\Bigg],
\end{eqnarray}
where and the upper/lower sign corresponds to the $u/d$-quarks.
The free energy $F$ is related to the thermodynamic potential 
$\Omega$ by the relation $F = \Omega +\mu_u\rho_u+\mu_d\rho_d$
and, as already discussed for non-relativistic superconductors,
the energy should be  minimized at constant temperature and 
density of the matter at various  flavor 
asymmetries defined as $\alpha \equiv (\rho_d-\rho_u)
/(\rho_d+\rho_u)$ \cite{DFS3}.  Eqs. (\ref{eq5a}) and (\ref{eq5b}) 
are the (ultra)relativistic counterparts of Eqs. (\ref{eq1a}) and
(\ref{eq1b}) which, apart from the relativistic form of the spectrum,
include the contribution of anti-particles.
\begin{figure}[tb] 
\begin{center} 
\includegraphics[angle=0,width=\linewidth]{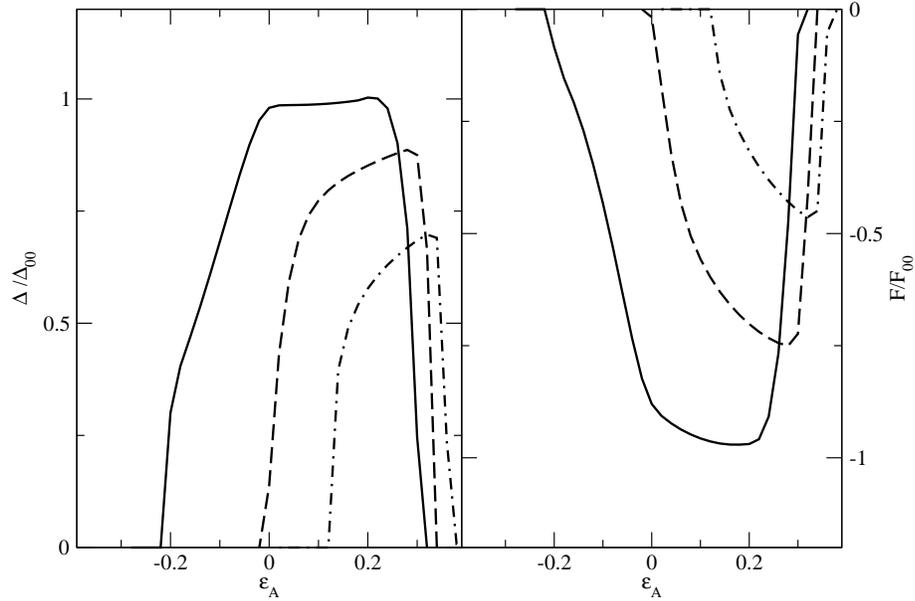}
\end{center}
\caption{
The color superconducting gap (left panel) and the 
free energy (right panel) as a function of 
relative deformation parameter $\varepsilon_A$ for 
the values of the flavor asymmetry $\alpha = 0$ (solid lines), 0.1
(dashed lines),  0.2 (short dashed lines) and 0.3 (dashed dotted
lines) at density $\rho_B = 0.31$ fm$^{-3}$ and temperature $T=2$
MeV \cite{DFS3}.}
\end{figure}

To obtain the selfconsistent solutions we employ a 
three-dimensional momentum space cut-off $\vert {\bf p}\vert <
\Lambda$ to regularize the divergent integrals. 
The phenomenological value of 
the coupling constant $G_1$ in the $\langle q q\rangle $  Cooper
channel is related  to the coupling constant in the $\langle q\bar q\rangle $ 
di-quark channel by the relation $G_1 = N_c/(2N_c-2) G$;
the latter coupling constant and the cut-off
are fixed by adjusting the model to the vacuum properties of the 
system  \cite{Schwarz:1999dj,Buballa:2001wh}.
Figure 9 summarizes the main features of
the color superconducting 
DFS phase \cite{DFS3}. The physically relevant regime of flavor 
asymmetries which is likely to occur in the charge-neutral matter 
under $\beta$ equilibrium  is $0.1 \le \alpha \le 0.3 $
\cite{Iida:2002ev,Steiner:2002gx,Blaschke,Huang:2003xd}.
The dependence of the color superconducting gap (left panel) and the 
free energy difference between the superconducting state 
and normal state (right panel) are shown as a function of 
deformation parameter $\varepsilon_A$ for several flavor asymmetries
at the baryonic density $\rho_B = 0.31$ fm$^{-3}$ and 
temperature $T=2$ MeV (the conformal deformation  $\varepsilon_S =0$).
The $\beta$ equilibrated quark matter requires an excess of the 
$d$ over $u$ quarks, therefore the range of the flavor asymmetry 
is restricted to the positive values. 
The deformation parameter $\varepsilon_A$ assumes both
positive and negative values. The main features seen in Fig. 9 are
consistent with  the results obtained for the non-relativistic 
superconductors: for a fixed $\alpha\neq 0$ and $\varepsilon_A >0$, 
the gap is larger in the DFS state than in the ordinary BCS 
state ($\varepsilon_A=0$), i.e.,  the 2SC phase is unstable towards 
deformation of the Fermi surfaces. Accordingly,  the minimum of the 
free energy  corresponds to the DFS state with  $\varepsilon_A\simeq 0.25$ 
and its position weakly depends on the value of $\alpha$.

\section{Concluding remarks}

The properties of the asymmetric superconductors have been an exciting
subject since the advent of the BCS theory of superconductivity more
than four decades ago. While the early studies were motivated by the 
effects of the paramagnetic impurities on the superconducting state 
 and the possible coexistence of the ferromagnetic
and superconducting phases in metallic superconductors, 
the recent  work on this subject has been motivated by the need
to understand the  nucleonic superfluids, the colored quark
superconductors and the dilute trapped atomic gases.

This mini-review focused on several non-BCS phases that 
may be featured by the asymmetric superconductors. The 
main points are summarized below:
\begin{itemize}
\item For small asymmetries, the superconducting state 
is homogeneous and the order parameter
preserves the space symmetries. For most of the systems 
of interest the number conservation should be implemented by 
solving equations for the gap function and the 
densities of species self-consistently. In such a scheme the 
physical quantities are single valued functions of the 
asymmetry and temperature, contrary to the double valued results obtained 
in the non-conserving schemes.
\item For large enough asymmetries the homogeneous 
state becomes unstable towards formation of 
either the LOFF phase - a superconducting state with nonzero
center-of-mass momentum  of the Cooper pairs,  or 
the DFS phase -  a superconducting state which requires
a quadrapole deformation of Fermi surfaces.
A combined treatment of these
phases in non-relativistic systems shows that while the 
LOFF phase corresponds to a local minimum, the DFS phase 
has energy lower that the LOFF phase. These phases 
break either the rotational, the  translational or both symmetries.

\item The temperature dependence of the pairing gap for 
the homogeneous, LOFF 
and DFS superconducting phases shows the phenomenon of 
re-entrance:  the superconducting state is revived at finite
temperatures. There are two critical temperatures for the 
phase transitions from the normal to  the superconducting state 
and back as the temperature is increased from zero to 
finite values. 

\item The color superconducting DFS-phase, which is treated in a
four-fermion contact interaction model, is preferred 
to the homogeneous 2SC state for asymmetries that are 
typical to matter under  $\beta$-equilibrium.

\end{itemize}

\section*{Acknowledgments} I would like to thank 
Umberto Lombardo, Herbert M\"{u}ther, Philippe Nozieres, 
Peter Schuck, and  Hans Schulze  for their contribution to the 
research reported here. This work was
supported by a grant provided by the SFB 382 of the DFG.

\end{document}